\documentclass[aps,prb,preprint,groupedaddress]{revtex4}

\usepackage{graphicx}
\newcommand{\be}{\begin{equation}}
\newcommand{\ee}{\end{equation}}

\begin{document}

\title{Universal anisotropic condensation transition of gases in nanotube 
bundles}

\author{Silvina M. Gatica, M. Mercedes Calbi and  Milton W. Cole}

\affiliation{Department of Physics, Pennsylvania State University, 
University Park, PA 16802}

\date{\today}

\begin{abstract}

Gases adsorbed within bundles of carbon nanotubes (inside of the nanotubes 
or in the interstitial channels between the tubes) exhibit a variety of phase 
transitions with the help of interactions between molecules in neighboring 
channels or tubes. Because the channels/tubes are widely separated, these 
transverse interactions are weaker than the (longitudinal) interactions within 
the 
same channel. The transition temperatures that result are therefore lower 
than those of typical two- or three-dimensional transitions of the same 
species of molecules. We discuss here the condensation transition of such 
a gas to form a liquid, expressing the transition behavior in universal 
form, where the reduced critical temperature T$_c^*$ is a universal function 
of the reduced transverse interaction.

\end{abstract}
\maketitle

\section{Introduction}

A growing literature of experiments, simulations and theory is devoted to 
the study of various gases adsorbed either inside or outside of bundles of 
carbon nanotubes \cite{peas,revaldo,boni,boro,rmp,vilches,hecond,
gubbins,karl,kostov1,kostov2}. On the inside, adsorption may occur either 
within the tubes or in the interstitial channels (ICs) between tubes. On 
the outside, the adsorption of small molecules is believed to start in the 
grooves between nanotubes and then cover the surface with a film of 
increasing coverage. One of the most tantalizing aspects of this research 
is the possibility of observing one-dimensional (1D) or quasi-1D 
matter. By the term ``quasi-1D'', we mean those phases for which the 
transverse interactions (between particles in adjacent 1D chains) play a 
non-negligible role in determining the 
phase behavior of systems that in some respect are 1D; in particular, the 
transverse interaction is {\em essential} to producing a genuine 
thermodynamic transition at nonzero temperature (T). In fact, all of these 
phases are 3D (since the particles of interest are highly constrained by 
the nanotube lattice), so there is some ambiguity in the choice of 
terminology.

Such quasi-1D phases can occur, in principle, within the tubes, within the 
ICs or in the grooves. Two kinds of transitions have been discussed 
theoretically: condensation \cite{peas,rmp,hecond,gubbins,kostov2} 
and crystallization \cite{carraro}. Both 
kinds differ from their counterparts in 3D matter because of the extreme 
anisotropy of the ordered phase and the low values of the predicted 
transition temperatures, which are even lower than those of monolayer 
films \cite{hecond}. The reason for these differences is that the 
transverse interaction is much weaker than the longitudinal interaction 
(that within a given channel).

This paper describes universal behavior of the gas-liquid condensation 
transition in this environment, allowing one to deduce results for any 
particular system that is characterized by a set of assumptions. The key 
assumption is that the dominant interaction energy is that involving 
particles within a given channel, while the transverse interaction is much 
weaker. The longitudinal problem consists of a 1D fluid of particles 
interacting with Lennard-Jones (LJ) interactions:

\be
V_l(r) = 4 \varepsilon \left[\left(\frac{\sigma}{r}\right)^{12}-\left(\frac{\sigma}{r}\right)^6\right]
\ee 

Here, as usual, $\varepsilon$ and $\sigma$ are the well-depth and 
hard-core diameter of the potential. The transverse interaction between 
molecules is written in the form known to describe long-range interactions 
in free space:

\be
V_t= - \frac{C_6}{r^6}
\ee

\noindent The coefficient $C_6$ could be taken here as the free-space 
interaction coefficient of the particles, although there is theoretical 
indication that screening by the nanotubes alters this 
interaction significantly \cite{kostov2} . One might assume, for 
example, that the same fundamental LJ interaction is responsible for the 
transverse and longitudinal interactions, in which case (from Eq. 1) 
$C_6$ assumes the value

\be
C_{6,LJ} = 4 \varepsilon \sigma^6
\ee

A key aspect of the present paper is that we obtain the transition 
behavior in universal terms, so that the appropriate interaction 
parameters can be used to deduce quantities applicable to a system of 
special interest.

The outline of this paper is the following. The next section evaluates 
the transition behavior with  two models. One describes the problem in 
terms of an anisotropic lattice gas model, which has been studied 
previously \cite{hecond}. The alternative model is a continuum description, 
in which case the equation of state is evaluated by treating the 
transverse interaction as a perturbation of the longitudinal problem 
\cite{peas}. Perhaps surprising is the finding that the predicted 
transition temperatures are rather similar. Section III comments further 
upon the solution and related problems.

\section {Model Calculations}

In a preliminary  work \cite{peas} we discussed C$_{60}$ molecules
 adsorbed inside the nanotubes, a 1D system. If the 
interaction between molecules in neighboring tubes is considered, there 
is a transverse interaction that, even if it is weaker than the 
intrachannel interaction, is enough  to drive a phase transition to a 
condensed phase. 

Smaller atoms or molecules adsorbed in  the interior  of the tubes may 
have rich structures, depending on their size. In  the case of  H$_2$ 
and inert gases, the atoms are adsorbed first on the wall of the tube 
forming a shell \cite{axial}, and after completion of the shell, for a 
higher chemical potential, the center of the tube starts to be filled.
When that occurs, the coverage of the shell is similar to that of a 
monolayer on graphite and the system can be seen as  a superposition of 
a 1D gas constrained to the center (axial phase) and a thermodynamically 
inert solid 
shell covering the wall.  

We find a qualitatively similar situation in the case of atoms in the 
interstitial channels. For example, as predicted in a previous work 
\cite{rmp}, small atoms or molecules like He, Ne or H$_2$ can be adsorbed 
in the interstitial channels of a bundle of nanotubes, forming a  1D gas, 
that would condense due to the interaction with atoms in other 
interstices of the bundle.

In either of the two cases, the purely  1D gas can be described by the 
equation of state for Lennard-Jones interacting particles,
\cite{p1d,footnote}

\be
\frac{1}{a}=\frac{\int_0^{\infty} dz \;e^{-\beta[V_l(z)+zP_{1D}]}}
{\int_0^{\infty} dz \;z\;e^{-\beta[V_l(z)+zP_{1D}]}}
\ee

\noindent that expresses the 1D density, $N/L=1/a$, in terms of the 1D line 
pressure $P_{1D}$ and $\beta=1/(k_B T)$. Here $z$ is the coordinate 
along the channel, or axis of the tube.

The interaction with molecules in neighboring channels or tubes can be 
considered in a perturbative way, so the change in the free energy is 
given by

\be 
\Delta F_{inter} = \frac{N \nu}{2 a} \int_{-\infty}^{\infty} 
\,dz \; V_t[(b^2 +z^2)^{1/2}]
\ee

Here $b$ is the separation between tubes or channels and $\nu$ is the 
number of nearest neighbor tubes or channels. For atoms inside the tubes, 
$b=17$ \AA$\,$ (for (10,10) tubes) and $\nu=6$, since the tubes form a 
hexagonal lattice, while for  IC atoms the lattice is honeycomb, with 
$\nu=3$ and $b=9.8$ \AA. In the case when $b$ is sufficiently large that 
the interaction may be approximated by the asymptotic form 
$V_t(r) \approx - C_6/r^6$, the integration yields 

\be
\Delta F_{inter} = - \nu N \; \frac{3 \pi}{16}\;\frac{C_6}{b^5}\;
\frac{1}{a} 
\ee

Including this term in the equation of state yields a relation for the 
shift of the 1D pressure proportional to $1/a^2$

\be
P(a,T)=P_{1D}(a,T) - \frac{\alpha}{a^2} \;\;\;\;; \;\;\alpha=\nu \,\frac{3 \pi}{16}\;
\frac{C_6}{b^5}
\ee

As a consequence, the pressure as a function of the density exhibits a van 
der Waals loop at low temperature, indicating the presence of a phase 
transition from a dilute phase to a higher density phase. The critical 
temperature for this transition  depends on the strength of the transverse 
interaction, $\alpha$. In Fig. 1 (solid line) we plot the scaled critical 
temperature deduced from Eq. 7

\be
T_c^* =  \frac{T_c}{\varepsilon}
\ee

\noindent as a function of the reduced transverse interaction strength, 

\be
\alpha^* = \frac{\alpha}{\varepsilon\sigma}.
\ee

Note that Fig. 1 is a universal curve, from which one can extract the 
critical temperature for any gas, adsorbed in either type of array of 
1D channels. All the key information about the gas and array 
(i.e. $\varepsilon$, $\sigma$, $\nu$ and $b$) is included in the universal 
constant $\alpha^*$. For example, for He in an array of IC, $\alpha^*$ is 
$7.5 \; 10^{-3}$ resulting $T_c^*=0.1$, while for He in the interior of 
the tubes,  $\alpha^*$ is $9.5 \;  10^{-4}$ and   $T_c^*=0.075$.

Next, we study  the system as an anisotropic  lattice gas. The sites of 
the lattice are separated by a typical interatomic distance $a$ along the 
channels, and the channels form an array (hexagonal or honeycomb for the 
interior sites or interstitial channels \cite{kostov1} respectively). Fisher 
proved that the critical temperature for an anisotropic Ising model 
in a cubic lattice has the form \cite{fisher}, 

\be
k_B T_c =\frac{2 J_l}{\ln(1/c)-\ln[\ln(1/c)]}
\ee

\noindent where $J_l$ is the interaction along the channel, and 
$c = J_t/J_l$, where $J_t$ is the transverse interaction. This formula is 
valid only in the very anisotropic case, i.e. $c<0.1$ and includes only  
nearest neighbors' interactions. To adapt this model to our system, we 
choose $J_l = V_l(a)/4$ (the  usual conversion from Ising model to a 
lattice gas), and

\be
J_t = \eta \frac{\bar V_t(b)}{4}
\ee    

\noindent where $\eta$ is the coordination number of the lattice (3/2 
and 3 for honeycomb or hexagonal, respectively) and  $\bar V_t = f V_t$ 
represents the interaction between an atom in one channel with $f$ 
atoms in a neighboring channel. We introduce the factor $f$ to account 
for the fact that if $b>>a$, we count the transverse interaction 
with more than one atom  in  the neighboring channel. Actually, $f$ is 
proportional to $b/a$ and is calculated from an equation
expressing the effective nearest neighbor interaction in terms of the net 
effect of all neighbors,

\be
\bar V_t = \sum_i V_t[(b^2+z^2_i)^{1/2}] 
\ee

\noindent where the sum is over all the atoms in neighboring channels, 
in positions $z_i$  along the channel. If $b>>a$, the sum can be 
approximated by the integral

\be
\frac{1}{a} \int_{-\infty}^{\infty} dz \, V_t[(b^2+z^2_i)^{1/2}]  = f V_t(b)
\ee 

\noindent with

\be
f = \frac{3\pi}{8} \frac{b}{a}
\ee

\noindent This factor, for example,  equals 4.5 in the case of He in the 
interstitial channels, with $a=\sigma$. With this consideration, the 
factor $c$ is identical  to $\alpha^*$.

In Fig. 1, dashed line, we plot the resulting critical temperature as a 
function of $c=\alpha^*$. Note the similarity of the trends predicted with 
the two different approaches, as previously found for the particular  
case of C$_{60}$ \cite{peas}. This occurs because this transition happens 
at a low value of T, when the longitudinal correlation length is so long 
that the (essentially mean field) perturbation theory is expected to work 
\cite{fisher2}.

In Table I, we display the critical temperatures for some gases adsorbed 
inside the tubes or in the interstitial channels, according to  
perturbation theory and the lattice gas model. SF$_6$ is included in the 
table but it is not clear whether it  would form a  1D phase inside the 
tubes, even as an axial phase.

\section{Summary}

   We have found, with alternative models, the critical temperatures for 
an anisotropic fluid 
within a nanotube bundle. The resulting expresion for T$_c^*$ exhibits a 
universal dependence  on the reduced interaction strength. Such 
transitions would be extremely interesting to find experimentally. 

The present model makes a number  of simplifying assumptions. The neglect 
of disorder (due to impurities and thermal vibrations) appears to be the 
most serious approximation. An aditional assumption is the use of free 
space pair interactions for the numerical results, omitting the effect of 
interaction screening by the tubes  and elastic-mediation of the 
interaction due to dilation and deformation of the tube lattice 
\cite{dilation}.

We are grateful to National Science Foundation for support of this work.

\newpage
 
\begin{table}
\begin{tabular}{|c|c|c|c|c|c|c|c|c|} \hline \hline
      && &\multicolumn{3}{|c|}{IC} & \multicolumn{3}{|c|}{NT}  \\ \hline
          &$\;\varepsilon$ (K)$\;$&$\;\sigma$ (\AA)$\;$&$\;\alpha_{IC}^*\;$&$\;
T_{c,\,pert}^*\;$&$\;T_{c,\,LG}^*\;$&$\;\alpha_{NT}^*\;$&$\;T_{c,\,pert}^*\;$&$\;T_{c,\,LG}^*\;$     \\ \hline
$\;\;$ He $\;\;$&10.2&2.6 &$\;0.0075\;$&0.10 & 0.15&$\;0.00095\;$& 0.075&0.098\\
 Ne& 35.6&2.8&0.0096&0.11 &0.16 &0.0012& 0.079&0.098\\
 H$_2$&37&3.1 &0.02 &0.12 &0.20 &0.0026 &0.089 & 0.12\\
 Ar&120&3.4&&&&0.0029&0.092 & 0.13\\
 Kr&171&3.6&&&&0.004&0.094&0.13\\
 Xe&221&4.1&&&&0.0066&0.10&0.15\\
 SF$_6$&208&5.3&&&&0.04&0.15&0.24\\
 C$_{60}$&2300&9.2& &&&0.1&0.20&0.34 
\\ \hline\hline
\end{tabular}
\caption{Reduced critical temperature $T_c^*=T_c/\varepsilon$ and reduced transverse interaction $\alpha^*=\alpha/(\varepsilon \sigma)$ for various gases adsorbed in the interstitial channels (IC) or inside the nanotubes (NT), calculated by perturbation theory ($T_{c,\,pert}^*$) or from the lattice gas model ($T_{c,\,LG}^*$). Values of $C_6$ entering in $\alpha^*$ were taken from Ref. 18 in the case of He, Ne, Ar, Kr and Xe. For the rest of the gases, a Lennard-Jones transverse interaction was asummed, with $C_6 = 4 \varepsilon \sigma^6$. The empty spaces in the table occur because large atoms or molecules are belived to be excluded from the ICs.}
\end{table}

\newpage

\begin{figure}
\includegraphics[height=4in]{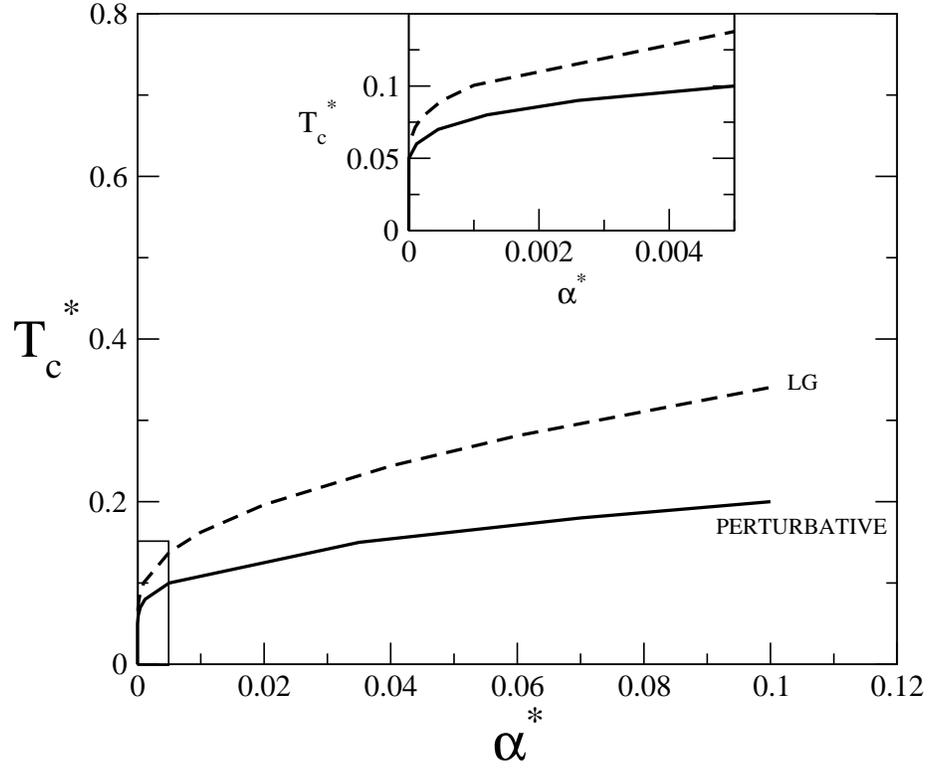}
\caption{Reduced critical temperature T/$\varepsilon$ as a function of the reduced 
strength $\alpha^*$, defined in the text, from a perturbative calculation (solid line) 
and lattice model (dashed line). Inset is an expanded version of the small 
$\alpha^*$ region.}
\end{figure}
 
\end{document}